# Topology-Driven Vibrations in a Chiral Polar Vortex Lattice

Eric R. Hoglund[1*∥], Harrison A. Walker[2,3∥], Peter B. Meisenheimer[4], Thomas W. Pfeifer[1,5], Niels De Vries[6], Dipanjan Chaudhuri[6], Ting-Ran Liu[7], Steven C. Quillin[8], Sandhya Susarla[9], De-Liang Bao[2], Patrick E. Hopkins[5,10,11], Andrew R. Lupini[1], Peter Abbamonte[6], Yu-Tsun Shao[7,12], Ramamoorthy Ramesh[4,13], Sokrates T. Pantelides[2,3,14¶], and Jordan A. Hachtel[1♠]

[1] Center for Nanophase Materials Sciences, Oak Ridge National Laboratory, Oak Ridge, TN 37830, USA

[2] Dept. of Physics and Astronomy, Vanderbilt University, Nashville, TN 37235, USA

[3] Interdisciplinary Materials Science Program, Vanderbilt University, Nashville, TN 37235

[4] Dept. of Materials Science and Engineering, University of California, Berkeley, CA 94720, USA

[5] Dept. of Mechanical and Aerospace Engineering, University of Virginia, Charlottesville, VA 22904, USA

[6] Materials Research Laboratory, University of Illinois, Urbana-Champaign, 104 S. Goodwin Ave., Urbana, IL, 61801, USA

[7] Mork Family Department of Chemical Engineering and Materials Science, University of Southern California, Los Angeles, CA 90089, USA.

[8] Bruker AXS LLC, 11511 NE 118[th] St, Kirkland, WA 98034, USA

[9] Materials Science and Engineering, School for Engineering of Matter, Transport, and Energy, Arizona State University, Tempe, AZ, USA

[10] Dept. of Materials Science and Engineering, University of Virginia, Charlottesville, VA 22904, USA

[11] Dept. of Physics, University of Virginia, Charlottesville, VA 22904, USA

[12] Core Center of Excellence in Nano Imaging, University of Southern California, Los Angeles, CA 90089, USA

[13] Dept. of Materials Science and Nanoengineering, Department of Physics and Astronomy, Rice University, Houston, TX, 77251 USA

[14] Dept. of Electrical and Computer Engineering, Vanderbilt University, Nashville TN 37235, USA

∥Contributed equally to the paper
*hoglunder@ornl.gov
¶pantelides@vanderbilt.edu
♠hachtelja@ornl.gov

# 1 Abstract

The ordering of magnetic or electric dipoles leading to real-space topological structures is at the forefront of materials research as their quantum mechanical nature often lends itself to emergent properties. Atomic lattice vibrations (phonons) are often a key contributor to the formation of long-range dipole textures based on ferroelectrics and impact the properties of the emergent phases. Here, using monochromated, momentum-resolved electron energy-loss spectroscopy ($q$EELS) with nanometer spatial resolution and meV-spectral-precision, we demonstrate that polar vortex lattices in $PbTiO_3$ spatially modulate the material's vibrational spectrum in patterns that directly reflect the overlying symmetry of the topological patterns. Moreover, by combining experiments with molecular dynamics simulations using machine learned potentials we reveal how these structures modify phonon modes across the vibrational spectrum. Beyond simple intensity modulation, we find that the chirality of the vortex topology imparts its unique symmetry onto phonons, producing a distinctive asymmetrical spectral shift across the vortex unit cell. Finally, the high spatial resolution of the technique enables topological defects to be probed directly, demonstrating a return to trivial $PbTiO_3$ modes at vortex dislocation cores. These findings establish a fundamental relationship between ferroelectric-ordering-induced topologies and phonon behavior, opening new avenues for engineering thermal transport, electron-phonon coupling, and other phonon-mediated properties in next-generation nanoscale devices.

## 2  Introduction

Polar topologies are analogous to magnetic topologies, but instead of the spins ordering, the spontaneous dipoles order. Since the discovery of polar vortices in $PbTiO_3$ in 2016[1] and polar skyrmions in 2019[2] there has been excitement due to their potential to host negative capacitance[3], their ability to have their optical properties controlled by electric fields[4], and their possible device applications.[5–8] In the case of polar vortices, unique collective atomic lattice vibrations (phonons) can emerge that are induced by the topology of the lattice, such as vortexons, in which the spiraling displacements of the polarization switch vorticity at GHz frequencies.[9] On contrary, the expected broad-spectrum redistribution of modes, including optic modes, has not yet been observed. Additionally, the lack of inversion symmetry in uniaxial ferroelectrics has been shown to result in chiral Weyl phonons that are switchable with the ferroelectric polarization and lead to a nonlinear phonon Hall effect.[10,11] In all these circumstances, the chirality of the vortex unit cell plays a dominant role on the emergent properties, by accentuating the effects of symmetry on the lattice vibrations.[12–14]

The length-scale of variations in the atomic-scale lattice within the vortex topologies necessitates high-resolution analysis. As a result, scanning transmission electron microscopy (STEM) plays a major role in their analysis due to the sub-Ångstrom spatial resolution of the technique.[1,2] Most experiments have focused on structural changes in the material based on atomic-resolution imaging or four-dimensional STEM (4D-STEM)[13–16]. However, the combination of spatial and spectral resolution available in electron energy-loss spectroscopy (EELS) within the STEM provides an intriguing opportunity to expand on these experiments by directly linking the polar topologies with localized phonon behavior.[17–22] In this technique, not only can nano/atomic-scale vibrational modes be directly probed[23–25], but also the scattering angle can be used to extract momentum and polarization information at this same scale.[26–28]

Here we directly measure the influence of the vortex topology in $PbTiO_3$ on the broad-spectrum atomic vibrations in real-space with monochromated electron energy-loss spectroscopy (EELS) in a STEM. The observed vibrational response demonstrates extreme asymmetry, with the peak of the vibrational response red-shifting along one domain wall, and blue-shifting along the other (or 'swooping') between vortex unit cells. Moreover, by examining the momentum-resolved EELS response, dichroic vibrational response from the Bragg peak Friedel pairs can be observed demonstrating the link of the asymmetric vibrations to the Friedel-symmetry breaking from the system's non-centrosymmetric dipolar arrangement. Furthermore, the response can be directly linked to the chirality of the system through molecular dynamics (MD) simulations using machine-learned (ML) interatomic potentials. Symmetry mapping with 4D-STEM and examining the vibrational response in reciprocal space using momentum-resolved EELS in reflection geometry (R-EELS),[22,29] combined with monochromated STEM-EELS further elucidate the connections between topological lattice periodicity, anisotropy, chirality, and defects alter unique collective lattice modes.

## 3 Results

Here we examine a $SrTiO_3$-$PbTiO_3$ heterostructure on $DyScO_3$, which is known to form a vortex lattice due to the interplay between electrostatics and elastic effects.[13,30–32] The substrate is then removed by chemical etching leaving a suspended thin film without the typical ion-mill surface damage, for observation in STEM.

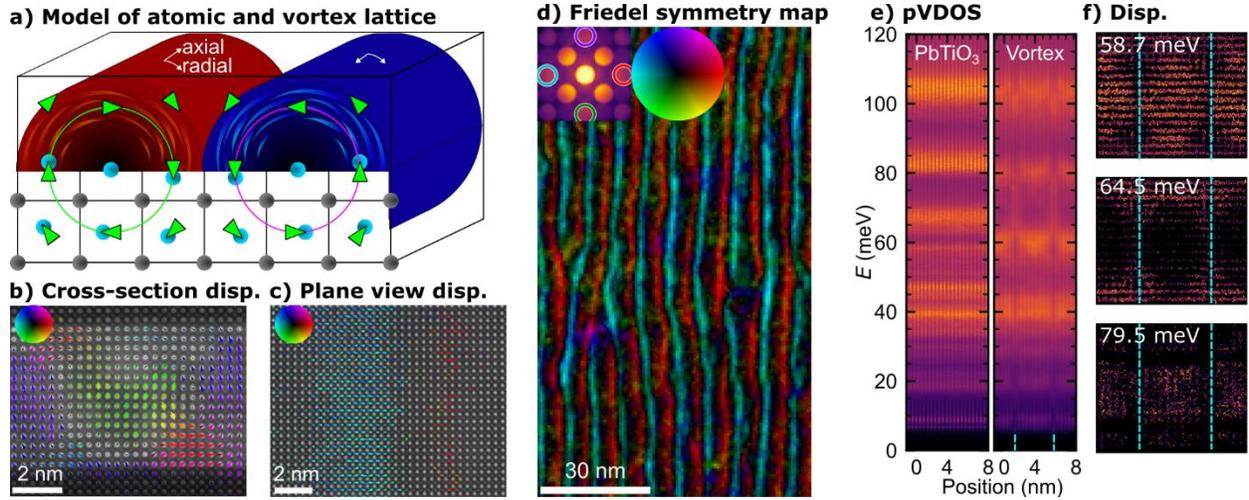

Figure 1. **Structure and Predicted Phonons of polar vortices.** (a) Model showing vortex lattice polarization. The top half represents the two-vortex basis with opposite axial polarization (white arrows) but same handedness, while the bottom demonstrates how the vortices are composed of an atomic lattice. HAADF image of a (b) cross-sectional and (c) plan-view samples with atomic displacements showing a vortex pair from front and top projections in (a). Displacements are mapped onto a complex color wheel. (d) 4D-STEM effective polarization map showing the signature ordering of the vortex lattice (polarization normal to vortex axis). The left inset in (d) shows the position averaged diffraction pattern with the $\pm g_{20}$ and $\pm g_{02}$ Friedel pairs (annotated) used as the basis for the center-of-mass signal that is projected onto a complex color wheel (right inset). (e) Spatially projected Vibrational Density of States (pVDOS) for uniaxial $PbTiO_3$ and Vortex along the axis perpendicular to the vortex tubes. Blue annotated lines indicate the center position of the vortices. (f) Frequency resolved displacements corresponding to 58.7, and 64.5, and 79.5 meV showcasing the interaction of vibrations with the exterior and interior of the vortices.

A model of a typical two-vortex basis in these samples is shown in Figure 1(a), where the lower half shows the Ti displacements (**u**) within $PbTiO_3$ atomic unit cells composing the vortices and the top half schematically shows the polarization of the cells comprising the two larger vortices of opposite winding around their axis. The winding can be characterized by the vorticity ($\mathbf{v} = \nabla \times \mathbf{u}$) and the helicity ($H = \mathbf{u} \cdot \mathbf{v}$). A high-resolution atomic-number contrast image of a cross-sectional sample (that was not lifted off from the substrate) is shown in Figure 1(b). It exhibits ~11 nm periodic displacements of the Ti sublattice around two vortex cores, as seen in the front face of the model. A high-resolution atomic-number contrast, plan-view image of a lifted-off sample (Figure 1(c)) shows the same ~11 nm periodic displacements of the Ti sublattice confirming the existence of a polar topology. We note that the electron probe is preferentially sensitive to the top-half of the sample such that the projected displacements are not averaged out from the top to the bottom of the vortices and that axial polarizations are small relative to the radial ones. This finding agrees with prior studies and explains why each vortex shows opposing polarization, like the top half of the model in Figure 1(a).[15] The preferential sensitivity to the top surface also suggests that the larger proportion of displacements to the left (cyan) relative to those to right (red) are from the vortices being offset along the film normal. This staggering is important for the structure and properties of the vortex lattice and is one of two sources of vortex lattice chirality.[33]

Using 4D-STEM we create an effective polarization map (Figure 1(d)) using the dichroic intensity of diffracted Friedel pairs (*i.e.* $\pm\boldsymbol{g}$ diffraction peaks shown in Figure S1(a)) resulting from dynamical diffraction effects reflecting a lack of inversion symmetry.[13–15] The detected polarization is perpendicular to the vortex axis and the probe, agreeing with the displacement maps in Figure 1(c), which originates from the radial polarization on the top half of the sample and confirms that the polar topology is the vortex phase rather than a skyrmion or meron labyrinth.[13,15] The preferential sensitivity to top half of the vortices in the plan-view sample is related to the electron channeling condition with specific probe convergence angles and has been explored elsewhere.[34] From the larger microns field-of-view image, we see that the vortices are ordered into a one-dimensional lattice, which also gives rise to satellite peaks in the parallel beam diffraction pattern (Figure S1b). It is worth noting that, for an equiaxial strain state, the equilibrium topological phase is a disordered arrangement of skyrmions. The vortices are also stable in the ML-MD simulations that include the strain relaxation but not the long-range electrostatic forces (see Methods), which could indicate that strain plays a dominate role over electrostatics. Retention of a vortex lattice after lifting off the sample and relaxing strain indicates that the vortex lattice persists as a metastable phase, which suggests routes for tuning the symmetry of topological phases that are different than the underlying atomic lattice.

In the larger field-of-view image we can additionally see the presence of edge dislocations in the vortex lattice that are a form of topological defect. The strain relaxation also results in vortex lattices forming 90° super-domains that we observe in real-space images and diffraction, which is visible as two sets of satellite peaks in Figure S1 and later in Figure 3(a). The presence of both domains and dislocations demonstrate a rich landscape of defects that can be observed with spatially resolved techniques.

To understand the influence of the vortex lattice on atomic vibrations, we performed machine learning molecular dynamics (ML-MD) simulations comparing two systems: a reference supercell of $PbTiO_3$ with uniaxial polarization and a $PbTiO_3/SrTiO_3$ superlattice containing the polar vortex lattice. The simulations employed DFT-based ML potentials to model the atomic interactions and calculate vibrations in supercells of up to 216,000 atoms which is well beyond the capabilities of DFT. Figure 1(e) displays a cropped region of the supercell (Figure S2) spatially resolved VDOS for both the atomic and vortex lattices, which reveals significant spectral shifts up to 5 meV induced by the topological polarization structure. While the uniaxial polarization produces a uniform vibrational response across the material (Figure 1(e) left) the vortex lattice induces a periodic softening of vibrations at the vortex cores (Figure 1(e) right). For example, we observe characteristic modifications in the optical phonon modes between 60-80 meV, where the vortex lattice leads to both frequency shifts and unique vibrational modes not existing in the uniaxial case. This periodic softening extends across the entire phonon spectrum, demonstrating how the topological structure fundamentally reorganizes the material's vibrational landscape.

We visualize the unique vortex vibrations in real-space by plotting amplitudes of atomic displacements oscillating with a given frequency, as shown in Figure 1(f). The magnitude of the displacements is modulated by the vortices. At 58.7 meV, the material exhibits enhanced displacements concentrated at the vortex cores, while at 65.3 meV the material shows stronger displacements in the regions where polarization vectors converge and diverge between adjacent vortices. These spatially resolved displacement patterns provide direct evidence that the polar vortex topology induces characteristic modulations in the atomic vibrations that reflect the symmetries of the overlying vortex lattice.

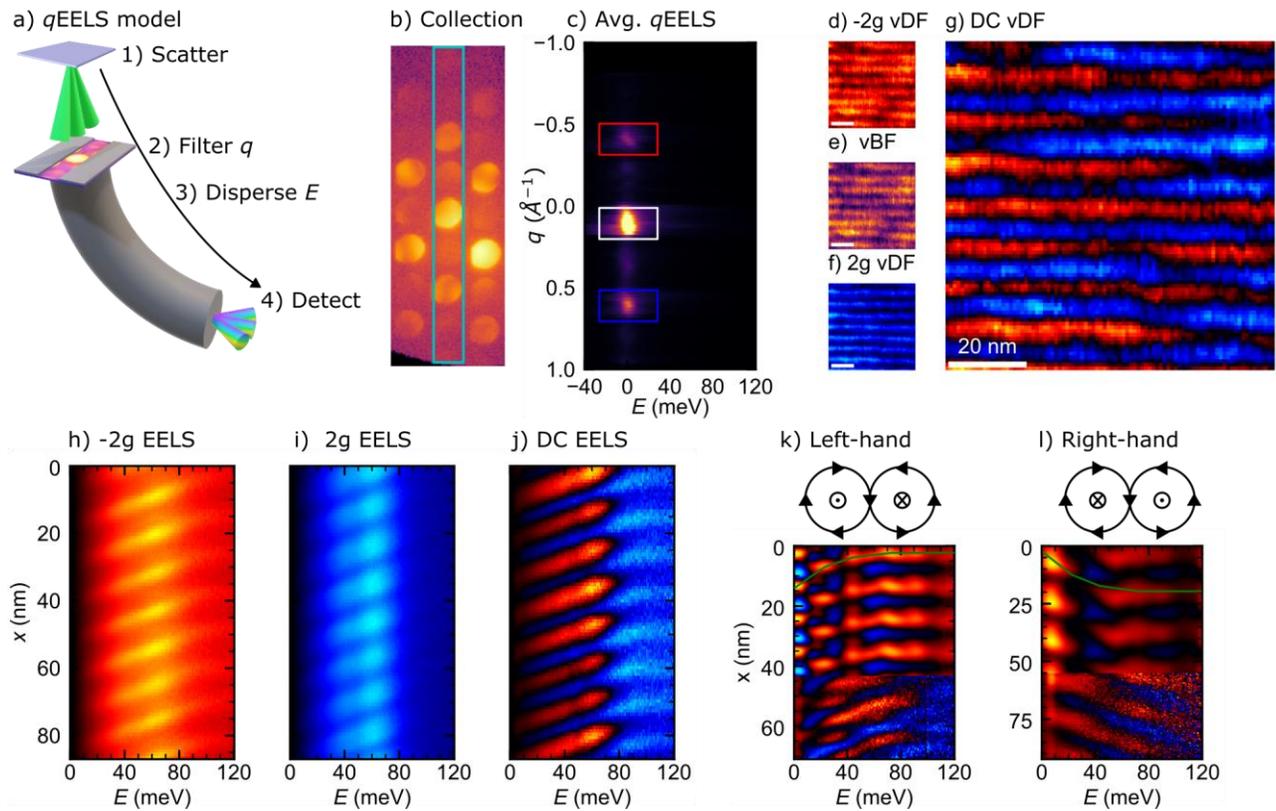

Figure 2. **Spatially resolved dichroic vibrational response.** (a) Schematic describing *q*EELS acquisition. (b) Diffraction pattern from the sample oriented so that $g_{H0}$ is aligned along the slot aperture (cyan), which is step (2) in (a). (c) Spatially averaged *q*EELS that is dispersed in energy and shows the filtered diffraction peaks at zero-energy, which is step (4) in (a). (d) Virtual bright-field (vBF), and (e) -2g and (f) 2g virtual dark-field (vDF) are reconstructed from the annotated zero-loss peaks in (c). (g) The zero-loss filtered vDF dichroic signal shows an effective

polarization similar to Figure 1(d). (h) -2$\boldsymbol{g}$ and (i) +2$\boldsymbol{g}$ vDF EELS profiles taken perpendicular to the vortices and their (j) dichroic signal showing spatially dependent dichroic vibrational loss at different regions following the vortex lattice periodicity. (k, l) Schematic (top) and simulated (bottom) vibrational EELS spectrum line scan across vortices with (k) left- and (l) right-handed chirality. The simulations are overlaid on two additional datasets (>40 nm), which shows a different swooping direction for left- and right-handed vortex unit cells. The full dataset in (k) and (l) are shown in Figure E1 and Figure 3, respectively.

To gain further insights into the lattice vibrations, we use $q$EELS in a STEM to map the spatial distribution of vibrations at discrete scattering angles as schematically shown for one probe position in Figure 2(a). This mode is like the 4D-STEM diffraction imaging performed earlier, where a two-dimensional scattering pattern was collected at each probe position on the sample, but instead a slot aperture is used to select a momentum path (Figure 2(b)) which the spectrometer then disperses as a function of $E$ (Figure 2(c)), thereby generating a multidimensional-STEM map as a function of position, momentum, and energy. An important aspect of the variable STEM optics is that there is a fundamental trade-off in spatial versus momentum resolution. Here the momentum resolution is set such that one atomic lattice Brillouin zone is integrated by choosing the probe semi-angle such that Bragg discs are just touching as shown in Figure 2(b). This condition provides spatial-resolution on the order of the atomic lattice, allowing for imaging of the vortices, while also maintaining Brillouin-zone-scale momentum resolution, which provides phonon eigenvector selectivity and reduces complications from interfering Bragg discs. By zero-loss energy-filtering the spectra we can create a vBF image (Figure 2(e)), vDF images (Figure 2(d,f)), and their dichroic signal from polarity-sensitive Bragg reflections (Figure 2(h)) to self-consistently create effective polarization maps, like in Figure 1(d), along a projected axis from the EELS data and directly image the polarization of individual vortices. From the dichroic vDF we see that the dislocation has a reversal of perpendicular polarization at the dislocation core, which means there is a local 1-dimensional reversal of helicity similar to the 2-dimentional reversal at some 0° domain walls.[14]

We can then look at the vibrations as a function of energy for different regions of the vortex lattice. The same axially-integrated spectra as functions of position along the vortex lattice direction at $q=\pm 2g_{H0}$ (Figure 2(h,i)) show that the spectral differences are periodic with the vortex lattice. The maximum response also continuously red shifts within a vortex unit cell, but is discontinuous between unit cells, creating an asymmetric "swoop" pattern rather than a symmetric sinusoid. By subtracting the $\pm 2g_{H0}$ profiles (Figure 2(j)) we find that the two spectra are dichroic, just like the effective polarization maps formed using the elastic scattering, but with the addition of a spectral dependence, indicating that new spatially periodic vibrations emerge from the long range order of the vortices.

We now consider the unexpected discontinuous swooping pattern and the induced symmetries of the vortex lattice. The vortex unit cell has a basis of two vortices with the same handedness but opposite axial components, which allows for two possible chiral unit cells, as schematically shown in Figure 2(k,l). Additionally, the offsetting of vortices seen in Figure 1(c) is another source of chirality. The response should nominally be symmetric as the polarization follows left-zero-right-zero polarization within a single vortex unit cell which is continuous with neighboring unit cells, which is observed in the sinusoidal effective polarization maps in Figure 1(d). Below the schematic of the left-hand vortex unit cell in Figure 2(k) is a simulated dichroic vibrational EELS line scan calculated from the ML-MD with another experimental dataset from a different vortex region shown below it (full dataset is shown in Figure E1). The observation of the spatially modulated vibrations is consistent with the spatial modulation in the spatially projected ML-MD and in good agreement with electron scattering simulations performed on the vortex structures under the same mixed real- and $q$-space conditions, confirming that the topology induces new collective lattice vibrations. Similarly, a right-handed simulation is shown Figure 2(l) with an experimental dataset shown below it (full dataset shown later in Figure 3(d)). We find that simulations match the "swooping" behavior of the experiments well and are mirrored with opposite slopes. We therefore demonstrate that this asymmetry is directly connected to the chirality of the vortex unit cell dictated by the topology in the system.

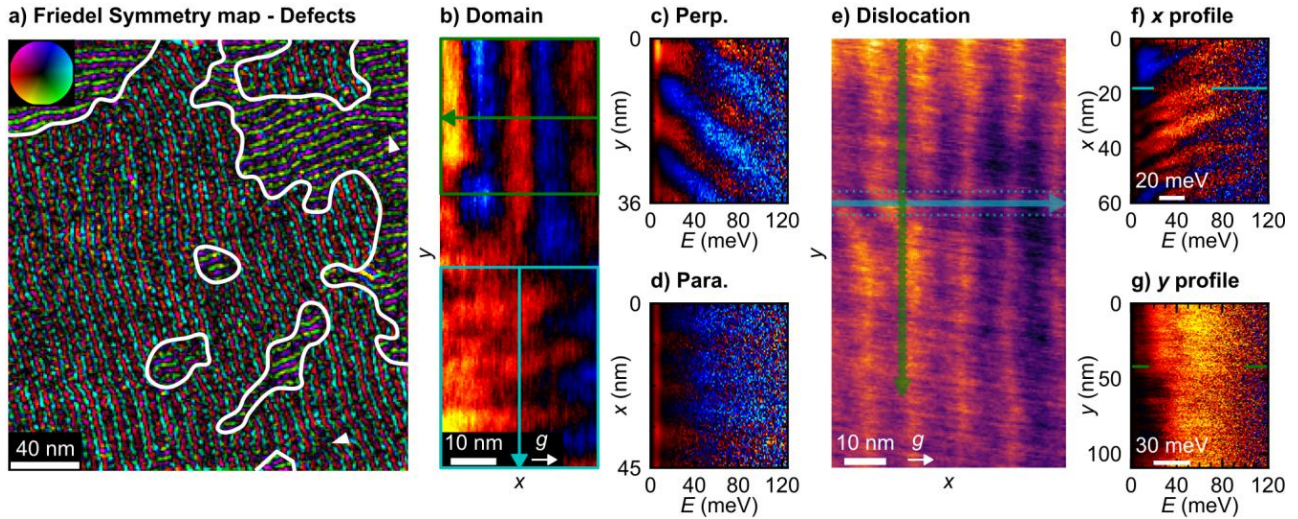

Figure 3. **Vortex lattice defects, asymmetry, and chirality impact of vibrations.** (a) Large field of view 4D-STEM effective polarization map showing domain boundaries (white outlines) and dislocations (white arrows). (b) Zero-loss filtered virtual dichroic signal from a region containing two perpendicular domains in an acquisition where the slot aperture is oriented along the horizontal axis ($x$), which is also the slow scan direction. In (b) annotate regions of integration mark domains perpendicular (green, slow scan axis) and parallel (cyan, fast scan axis) to the $2\boldsymbol{g}$ Bragg peaks selected by the slot aperture, with their respective position versus $E$ profiles shown in (d,e). (f) Enlarged region of the dislocation region from the zero-loss filtered BF image in Figure E1(c-f). EELS profiles (g) perpendicular and (h) parallel to the half-plane are annotated in (f) and their intersects at the dislocation core are annotated in (g,h).

We have so far shown that the polar topological lattice imparts its periodicity to the atomic vibrations, while focusing on the uniform regions. However, nearly all naturally occurring lattices contain defects (like dislocations) as well as 90º rotated super-domains in close proximity. Through the high spatial resolution of this technique the behavior in these regions can be probed directly. A region of the sample containing two 90º vortex domains is shown in Figure 3(b) allowing us to directly probe the relationship between the vortex lattice anisotropy on the vibrations, and the scattering directions selected by the qEELS slot relative to the vortex lattice. For the top domain the scattering is collected perpendicular to the vortex axis (see Figure S3), meaning in the same direction as radial polarization vectors of the vortices (as shown in Figure 1(d)). Here, we see an identical asymmetrical, dichroic, swoop response, as was observed in the more uniform region analyzed in Figure 2. However, for the bottom domain the scattering is collected parallel to the vortex axes, and hence normal to the radial polarization in the vortex itself. Here, the swooping spectral response is not observed at all, and moreover, the dichroic response is reduced to a small intensity modulation (Figure 3(f)). This comparison demonstrates that the signature swooping spectral shifts originate from the formation of the vortex lattice and that the anisotropy of the polar topology imparts its symmetry onto the collective vibrational behavior.

Figure 3(f) shows an enlarged view of the vortex lattice dislocation in Figure E1(c-f) with annotated profiles indicating where the dichroic vDF EELS in Figure 3(f, g) are obtained. From the dichroic vDF in Figure E1(f) we see that the dislocation has a reversal of perpendicular polarization at the dislocation core, which means there is a local 1-dimensional reversal of helicity similar to the 2-dimentional reversal at some 0º domain walls.[14] In the dichroic vDF EELS taken perpendicular to the vortices (Figure 3(g)) we see that the periodicity becomes less defined at the dislocation core, and a further comparison in Figure E1 shows that the swoop from the extra vortex is missing at the core. This feature shows that the unique vibrational modes coming from the vortex lattice are lost as the vortex helical polarization transitions through zero to the opposite helicity on the opposite side of the core. Similarly, a profile along the extra vortex Figure 3(g) shows a ~30 and 70 meV peak that converges into a single ~50 meV peak as the helicity flips. We therefore demonstrate the impact of the local dipole order on the vibrations and spatially resolve the impact of a topological lattice defect on the lattice vibrations.

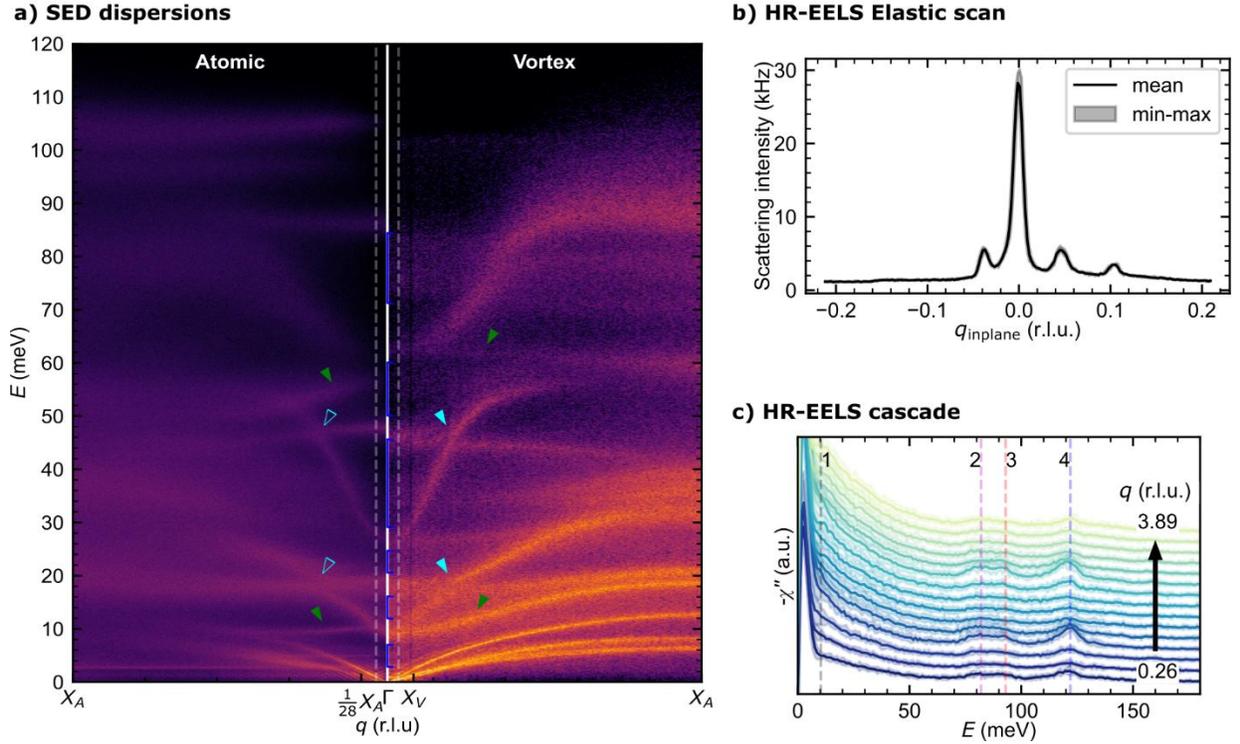

Figure 4. **Momentum-resolved vibrational response.** (a) SED dispersions from the (left) uniaxially polarized atomic lattice and (right) vortex lattice out to the atomic lattice Brillouin zone perpendicular to the vortices ($X_A$). The vortex Brillouin zone ($X_V$) and the equivalent fractional $X_A$ are also labeled for reference. Two band crossings (cyan) and two bands with large group velocity change (green) are annotated. Spectral weight along $X_V$ that is unique to the vortex cell are annotated in blue. (b) R-EELS Elastic scan showing Bragg scattering from the vortices. (c) Imaginary part of the charge electronic susceptibility ($\chi$) extracted from R-EELS along $X_V$ displayed as a $q^2$ normalized cascade that shows multiple orders of vortex Bragg peaks and four prominent loss peaks.

The observed broad-spectrum vortex lattice vibrations are periodic, indicating that there should be new phonon bands within a newly defined vortex Brillouin zone. Based on the vortex cell size, 15 atomic lattice branches ($3 \times 5$ atoms) could potentially have states redistributed to 3000 branches ($3 \times 600$ atomic cells $\times 5$ atoms). Figure 4(a) shows the phonon dispersion for both the uniaxially and vortex polarized cells extracted from the ML-MD using spectral energy density (SED), which allows for the simulation of the large unit cells necessary for the vortices and heterostructure while maintaining an accurate potential. The first thing to note is that the two dispersions share a very similar structure across the length-scale of the atomic lattice Brillouin zone with the vortex dispersion having a redistribution of spectral weight between marginally modified atomic lattice phonon bands. There is also an increased spectral weight within the newly formed Brillouin zone as indicated by the blue spans. The similar structure to $PbTiO_3$ with the addition of spectral weight within the vortex Brillouin zone implies to a first approximation that the vortex phonon dispersion can be considered as a superposition of atomic and vortex vibrations with the vortex vibrations originating from the redistribution of vibrational states from the emergent topology. Although the general structure of dispersions is similar, spectral shifts from softening of bands are observed, new bands emerge, forbidden crossings lead to a loss of intensity where bands meet (cyan arrows), and some bands slope that determine group velocities change significantly (green arrows). The clearest presence of new bands is one ~70 meV band extending across the full atomic lattice Brillouin zone and a splitting of one ~50 meV band prior to the annotated crossing into two prior to the crossing. These findings in reciprocal space demonstrate that the new emergent modes are collective at the length-scale of multiple vortex unit cells.

To experimentally measure changes in reciprocal space with high momentum resolution we use R-EELS (Figure 4(b,c)). The elastic scan Figure 4(b) shows periodic diffraction from the vortex lattice similar to T-EELS (Figure S1), which demonstrates that the reflection geometry is sampling the vortices in the $PbTiO_3$ layer. The $q$, $E$ cascade (Figure 4(c)) shows an intensity modulation following the elastic Bragg peaks, corresponding to the periodicity of the vortex lattice. Another straightforward observation is the presence of four peaks at finite energy loss, corresponding to phonon modes that modulate at a periodicity of the vortex Bragg reflections. Upon closer examination, the energy of the peaks is found to vary periodically (Figure S5), which confirms that dispersive phonons exist at these optical frequencies with spatial frequencies on the order of the vortex lattice. This momentum-space modulation at the length-scale of the vortex Brillouin zone complements the real-space T-EELS findings and furthermore experimentally demonstrates that the ordering of the vortex topology leads to a modification of the materials phonon dispersion.

## 4 Conclusions

Using a balance between spatial and momentum resolution in STEM-EELS we have shown that a polar topological vortex lattice impacts collective lattice vibrations across a broad spectrum by imparting its periodicity, anisotropy, and (chiral) symmetry. The vibrational properties depend on a superposition of the topological-lattice and the underlying atomic lattice. From a broader perspective, these results indicate a pathway for more tunable material's behavior as both the topology and the atomic lattice can be chosen, with the combination of topological- and atomic lattices allowing for more potential combinations than either alone. In addition, we find that defects in polar topology result in local vibrational behavior. Such defects are mobile and can be moved with fields or temperature, which may present a means to dynamically move local vibrational behavior throughout a sample.[35] The observed influence of vibrational modes by the overlying topological symmetry presents an opportunity to engineer phononic properties of materials, which in the case of these polar vortices could be switched through the application of external stimuli, such as electric fields that switch topology.

# 5 Materials and Methods

## 5.1 Samples

All samples were grown using reflection high-energy electron diffraction (RHEED) assisted pulsed laser deposition. RHEED was used during the deposition to ensure the maintenance of a layer-by-layer growth mode for both the $PbTiO_3$ and $SrTiO_3$. The specular RHEED spot was used to monitor the RHEED oscillations

To get a pristine plane view sample a sample of $(SrTiO_3)_{20}/(PbTiO_3)_{20}/(SrTiO_3)_{20}$ was grown on $DyScO_3$ substrate with a 16 nm $Sr_2CaAl_2O_6$ sacrificial layer and a 2.4 nm $SrTiO_3$ capping layer. For transfer, the heterostructure was spin-coated with a polymer support of 500 nm thick polymethyl methacrylate (PMMA) film and placed in deionized water at room temperature until the sacrificial layer was fully dissolved. The PMMA coated film was then released from the substrate and transferred onto the Si TEM chip. Finally, the PMMA layer was dissolved and removed from the membrane in acetone.

Cross-sectional $(SrTiO_3)_{16}/(PbTiO_3)_{20}/(SrTiO_3)_{16}$ samples were grown on $DyScO_3$ and then a Thermo Fisher Helios dual beam gallium focused io-beam was used to thin the samples. Initial energies were 30 kV, which was gradually decreased to 1 kV as the sample was thinned.

## 5.2 STEM Imaging and 4D-STEM

STEM imaging of the plane view sample shown in Figure 1(c) was acquired at 100 kV using a Nion UltraSTEM 100 with a 30 mrad convergence semi-angle. The cross-section image in Figure 1(b) was acquired using a Themis Z S/TEM operating at 200 kV with a 30 mrad convergence semi-angle. The 4D-STEM in Figure 3(a) was acquired at 100 kV using a Nion UltraSTEM 100 with a 5 mrad convergence semi-angle. 4D-STEM results shown in Figure 1(d) was acquired using a Thermo Fisher Scientific Spectra 200 S/TEM operated at 200 kV, with a probe semi-convergence angle of 2.3 mrad and a current of 100 pA. The 4D-STEM dataset was captured using an electron microscopy pixel array detector (EMPAD) over a 2D grid (256×256 scan) of real-space probe positions and an exposure time of 1 ms per frame.

## 5.3 Momentum Resolved Vibrational EELS

Vibrational EELS spectra were acquired at 60 keV using a Nion HERMES monochromated aberration-corrected dedicated STEM. A convergence angle of 5 mrad was set such that the atomic lattice Bragg discs are touching. This condition sets uncertainty such that spectra are Brillouin zone integrate in momentum-space while preserving atomic lattice scale spatial resolution (in the absence of aberrations). A 125 × 2000 µm entrance aperture was used at a 20 mm camera length such that the convergence and collection angles are similar, as shown in Figure S4. The initial direction of Bragg reflection was aligned using a defocused probe where the vortices could be seen parallel or perpendicular to each atomic Bragg reflection. However, upon refocusing the sample the main metric to determine the vortex direction relative to the Bragg reflections was consistent zero-loss filtered EELS images being consistent with the observed radial polarization observed in the elastic 4D-STEM experiments. Each spectrum along $x$, $y$, and $q$ was aligned using the center-of-mass.

## 5.4 ML-MD

Molecular dynamics (MD) simulations were carried out in LAMMPS using a DeepMD-based machine learning potential. The se_e2_a descriptor[36] was employed with a smoothing radius of 6 Å and a global cutoff of 10 Å. The descriptor network comprised hidden layers of 25, 50, and 100 neurons, while the fitting network consisted of three layers of 240 neurons each, all using tanh activation functions. The reference data consisted of density functional theory (DFT) calculations run in VASP under the Local Density Approximation (LDA) for $PbTiO_3$, $SrTiO_3$ (STO), and $PbTiO_3$/STO superlattice configurations generated by running molecular dynamics in the NVT ensemble at 200K, 300K, and 400K. Each calculation employed a plane-wave cutoff energy of 500 eV, generating a total of 21,127 training configurations, 5283 validation configurations, and 7475 test configurations. Training data was added iteratively until error was reduced to an acceptable level. While there is potentially a large overlap of configurations within this training set, we believe that it covers all relevant regions of phase space and is representative of the simulations performed in this work. The potential's performance was validated against the test configurations, achieving RMSE values of 0.279 meV/atom for total energy and 0.0493 eV/Å for forces on $PbTiO_3/SrTiO_3$ superlattice configurations, 0.225 meV/atom and 0.0428 eV/Å on bulk $PbTiO_3$, and 0.352 meV/atom and 0.0396 eV/Å on bulk $SrTiO_3$. Detailed validation plots can be found in the Supplementary Information.

The $PbTiO_3/SrTiO_3$ superlattice was constructed using the Atomic Simulation Environment (ASE)[37]. Through grid search of lattice parameters, we determined that a configuration of 3.88 Å × 3.88 Å × 4.1 Å provided conditions for vortex stabilization in the NVT ensemble at 300K. Given the metastable nature of the polar vortex state in the absence of the $DyScO_3$ substrate, initialization of the polarization was performed by strategic displacement of titanium atoms to establish the desired vortex chirality. A comparative analysis of initialized versus non-initialized structures is presented in the Supplementary Information and it is shown that for the specified lattice parameters, we get spontaneous formation of the vortices. The structures simulated consisted of $(PbTiO_3)_{16}/(SrTiO_3)_{20}$ and $(PbTiO_3)_{10}/(SrTiO_3)_6$ superlattice configurations. The first structure focused on the cross-sectional view, only consisting of 15,120 atoms and 1 pair of vortices, with total dimensions 28 x 3 x 36. The final model was created for the plane view vibrational EELS simulations, containing 216K atoms and 3 pairs of vortices, with total dimensions of 60 x 45 x 16 unit cells. An additional simulation of 50 vortices, with dimensions 1400 x 3 x 36 and 756,000 atoms was performed to look at vibrational modes along the x direction with high momentum resolution.

We conducted molecular dynamics simulations on three system configurations: one containing a single vortex pair (2 vortices, $(PbTiO_3)_{16}/(SrTiO_3)_{20}$, one with (50 vortices, $(PbTiO_3)_{16}/(SrTiO_3)_{20}$ and another with three vortex pairs (6 vortices, $(PbTiO_3)_{10}/(SrTiO_3)_6$). All simulations were performed in the NVT ensemble at 300K utilizing the Nose-Hoover thermostat. Each system underwent a 250 ps equilibration period at 300 K followed by production runs of 100 ps in which the final 5 ps was recorded, utilizing a 5 fs timestep throughout.

The production run trajectories were analyzed using custom Python scripts to calculate vibrational density of states, spectral energy density, displacement maps, and EELS spectra. The methodologies utilized are described in the following paragraphs.

The vibrational density of states was calculated by means of a Fourier transform of the velocity autocorrelation function in the Green-Kubo formalism:

$$g(\omega) = \frac{1}{2\pi k_B T} \int_{-\infty}^{\infty} \langle \boldsymbol{v}(\boldsymbol{0}) \cdot \boldsymbol{v}(t) \rangle e^{-i\omega t} dt$$

where $g(\omega)$ is the vibrational density of states, $\boldsymbol{v}(t)$ is the velocity vector at time t, $k_B$ is the Boltzmann constant, and T is the temperature. By performing this calculation on a per-atom basis, we can view the vDOS in real space, allowing us to observe the periodic modulation of phonon states induced by the vortices.

To evaluate the dispersion relations of disordered structures of this size, spectral energy density (SED) was calculated utilizing custom python code following the formalism of coherent SED presented in our work on selection rules of vibrational EELS[38].

Displacement maps were calculated by time averaging the positions of atoms, taking care to consider periodic boundary conditions. The average position of each atom was then subtracted from the instantaneous position of each atom, resulting in displacement vectors. By taking a Fourier transform along the time axis, we can then utilize a gaussian band pass filter and an inverse Fourier transform to arrive at frequency filtered displacements, allowing us to view vibrations of a given frequency in real space.

The EELS simulations were performed using custom python code, which implements the TACAW methodology developed by Castellanos-Reyes, Zeiger, and Rusz.[39] EELS simulations in this work were performed using the TACAW formalism. Parameters of note include a convergence angle of 5 mrad, a beam energy of 60 keV, and 500 MD configurations per probe position. The simulation size provides a momentum space resolution of $0.027 \text{ Å}^{-1}$ and $0.036 \text{ Å}^{-1}$ in the x and y directions, respectively, allowing sampling of 638 k-points within the first Brillouin zone. The frequency resolution determined by the 5 ps total sampling time (500 MD configurations x 0.01 ps), is 0.8 meV. Simulations were performed on two vortex superlattice structures of opposite chirality. All parameters of these superlattices were identical, save for the direction of the vortex chirality. Collection apertures were placed on the 2G and -2G Bragg discs of the simulation, enabling off-axis signal to be computed.

## 5.5    Reflection EELS

The R-EELS setup is based on a modified commercially available aberration-corrected HR-EELS setup retrofitted with an eucentric three-circle goniometer, extensively described in previous work.[40] Lens and monochromator voltages were optimized in energy loss- and angle-space, to achieve an optimum in combined energy- and momentum resolution. The experimental energy- and momentum resolutions were estimated from the direct beam to be ~6 meV and 0.02Å⁻¹, respectively, and are in agreement with the energy width of the elastic line and the angle width of the specular reflection.

The thin films/samples were mounted as-grown on a Molybdenum puck using Tantalum strips. The sample was annealed at 650C for 10 hours in UHV, after which it was transferred to the spectrometer chamber, which has a base pressure of $5 \cdot 10^{-11}$ torr. The samples were optically aligned such that the scattering plane of the spectrometer was along $X_V$ axis and energy-loss spectra were obtained for different momentum transfer. Elastic momentum scans were monitored throughout to ensure the absence of surface degradation and beam instabilities. The obtained spectra were divided by the matrix elements to obtain S(q,ω) and then averaged. Dividing S(q,ω) by the Bose factor yields χ"(q,ω).[40]

# 6  References

[1] A. K. Yadav, C. T. Nelson, S. L. Hsu, Z. Hong, J. D. Clarkson, C. M. Schlepütz, A. R. Damodaran, P. Shafer, E. Arenholz, L. R. Dedon, D. Chen, A. Vishwanath, A. M. Minor, L. Q. Chen, J. F. Scott, L. W. Martin, R. Ramesh, *Nature* **2016**, *530*, 198.

[2] S. Das, Y. L. Tang, Z. Hong, M. A. P. Gonçalves, M. R. McCarter, C. Klewe, K. X. Nguyen, F. Gómez-Ortiz, P. Shafer, E. Arenholz, V. A. Stoica, S.-L. Hsu, B. Wang, C. Ophus, J. F. Liu, C. T. Nelson, S. Saremi, B. Prasad, A. B. Mei, D. G. Schlom, J. Íñiguez, P. García-Fernández, D. A. Muller, L. Q. Chen, J. Junquera, L. W. Martin, R. Ramesh, *Nature* **2019**, *568*, 368.

[3] A. K. Yadav, K. X. Nguyen, Z. Hong, P. García-Fernández, P. Aguado-Puente, C. T. Nelson, S. Das, B. Prasad, D. Kwon, S. Cheema, A. I. Khan, C. Hu, J. Íñiguez, J. Junquera, L.-Q. Chen, D. A. Muller, R. Ramesh, S. Salahuddin, *Nature* **2019**, *565*, 468.

[4] A. R. Damodaran, J. D. Clarkson, Z. Hong, H. Liu, A. K. Yadav, C. T. Nelson, S.-L. Hsu, M. R. McCarter, K.-D. Park, V. Kravtsov, A. Farhan, Y. Dong, Z. Cai, H. Zhou, P. Aguado-Puente, P. García-Fernández, J. Íñiguez, J. Junquera, A. Scholl, M. B. Raschke, L.-Q. Chen, D. D. Fong, R. Ramesh, L. W. Martin, *Nat. Mater.* **2017**, *16*, 1003.

[5] Y. J. Wang, Y. L. Tang, Y. L. Zhu, X. L. Ma, *Acta Materialia* **2023**, *243*, 118485.

[6] X. Guo, L. Zhou, B. Roul, Y. Wu, Y. Huang, S. Das, Z. Hong, *Small Methods* **2022**, 2200486.

[7] J. Junquera, Y. Nahas, S. Prokhorenko, L. Bellaiche, J. Íñiguez, D. G. Schlom, L.-Q. Chen, S. Salahuddin, D. A. Muller, L. W. Martin, R. Ramesh, *Rev. Mod. Phys.* **2023**, *95*, 025001.

[8] V. A. Stoica, N. Laanait, C. Dai, Z. Hong, Y. Yuan, Z. Zhang, S. Lei, M. R. McCarter, A. Yadav, A. R. Damodaran, S. Das, G. A. Stone, J. Karapetrova, D. A. Walko, X. Zhang, L. W. Martin, R. Ramesh, L.-Q. Chen, H. Wen, V. Gopalan, J. W. Freeland, *Nat. Mater.* **2019**, *18*, 377.

[9] Q. Li, V. A. Stoica, M. Paściak, Y. Zhu, Y. Yuan, T. Yang, M. R. McCarter, S. Das, A. K. Yadav, S. Park, C. Dai, H. J. Lee, Y. Ahn, S. D. Marks, S. Yu, C. Kadlec, T. Sato, M. C. Hoffmann, M. Chollet, M. E. Kozina, S. Nelson, D. Zhu, D. A. Walko, A. M. Lindenberg, P. G. Evans, L.-Q. Chen, R. Ramesh, L. W. Martin, V. Gopalan, J. W. Freeland, J. Hlinka, H. Wen, *Nature* **2021**, *592*, 376.

[10] J. Im, C. H. Kim, H. Jin, *Nano Lett.* **2022**, *22*, 8281.

[11] W. Luo, J. Ji, P. Chen, Y. Xu, L. Zhang, H. Xiang, L. Bellaiche, *Phys. Rev. B* **2023**, *107*, L241107.

[12] P. Shafer, P. García-Fernández, P. Aguado-Puente, A. R. Damodaran, A. K. Yadav, C. T. Nelson, S.-L. Hsu, J. C. Wojdeł, J. Íñiguez, L. W. Martin, E. Arenholz, J. Junquera, R. Ramesh, *Proc Natl Acad Sci USA* **2018**, *115*, 915.

[13] Y.-T. Shao, S. Das, Z. Hong, R. Xu, S. Chandrika, F. Gómez-Ortiz, P. García-Fernández, L.-Q. Chen, H. Y. Hwang, J. Junquera, L. W. Martin, R. Ramesh, D. A. Muller, *Nat Commun* **2023**, *14*, 1355.

[14] S. Susarla, S. Hsu, F. Gómez-Ortiz, P. García-Fernández, B. H. Savitzky, S. Das, P. Behera, J. Junquera, P. Ercius, R. Ramesh, C. Ophus, *Nat Commun* **2023**, *14*, 4465.

[15] K. X. Nguyen, Y. Jiang, M. C. Cao, P. Purohit, A. K. Yadav, P. García-Fernández, M. W. Tate, C. S. Chang, P. Aguado-Puente, J. Íñiguez, F. Gomez-Ortiz, S. M. Gruner, J. Junquera, L. W. Martin, R. Ramesh, D. A. Muller, *Phys. Rev. B* **2023**, *107*, 205419.

[16] H. Ni, W. R. Meier, H. Miao, A. F. May, B. C. Sales, J.-M. Zuo, M. Chi, *Phys. Rev. Materials* **2024**, *8*, 104414.

[17] E. R. Hoglund, D.-L. Bao, A. O'Hara, S. Makarem, Z. T. Piontkowski, J. R. Matson, A. K. Yadav, R. C. Haislmaier, R. Engel-Herbert, J. F. Ihlefeld, J. Ravichandran, R. Ramesh, J. D. Caldwell, T. E. Beechem, J. A. Tomko, J. A. Hachtel, S. T. Pantelides, P. E. Hopkins, J. M. Howe, *Nature* **2022**, *601*, 556.

[18] O. L. Krivanek, T. C. Lovejoy, N. Dellby, T. Aoki, R. W. Carpenter, P. Rez, E. Soignard, J. Zhu, P. E. Batson, M. J. Lagos, R. F. Egerton, P. A. Crozier, *Nature* **2014**, *514*, 209.

[19] L. Haas, *Review of Education, Pedagogy, and Cultural Studies* **1995**, *17*, 1.

[20] Y.-H. Li, M. Wu, R.-S. Qi, N. Li, Y.-W. Sun, C.-L. Shi, X.-T. Zhu, J.-D. Guo, D.-P. Yu, P. Gao, *Chinese Phys. Lett.* **2019**, *36*, 026801.

[21] F. S. Hage, G. Radtke, D. M. Kepaptsoglou, M. Lazzeri, Q. M. Ramasse, *Science* **2020**, *367*, 1124.

[22] H. Yang, Y. Zhou, G. Miao, J. Rusz, X. Yan, F. Guzman, X. Xu, X. Xu, T. Aoki, P. Zeiger, X. Zhu, W. Wang, J. Guo, R. Wu, X. Pan, *Nature* **2024**, *635*, 332.

[23] E. R. Hoglund, D. Bao, A. O'Hara, T. W. Pfeifer, M. S. B. Hoque, S. Makarem, J. M. Howe, S. T. Pantelides, P. E. Hopkins, J. A. Hachtel, *Adv. Mater.* **2023**, *35*, 2208920.

[24] M. Xu, D.-L. Bao, A. Li, M. Gao, D. Meng, A. Li, S. Du, G. Su, S. J. Pennycook, S. T. Pantelides, W. Zhou, *Nat. Mater.* **2023**, *22*, 612.

[25] K. Venkatraman, B. D. A. Levin, K. March, P. Rez, P. A. Crozier, *Nat. Phys.* **2019**, *15*, 1237.

[26] F. S. Hage, R. J. Nicholls, J. R. Yates, D. G. McCulloch, T. C. Lovejoy, N. Dellby, O. L. Krivanek, K. Refson, Q. M. Ramasse, *Sci. Adv.* **2018**, *4*, 7495.

[27] R. Qi, N. Li, J. Du, R. Shi, Y. Huang, X. Yang, L. Liu, Z. Xu, Q. Dai, D. Yu, P. Gao, *Nat Commun* **2021**, *12*, 1179.

[28] C. A. Gadre, X. Yan, Q. Song, J. Li, L. Gu, H. Huyan, T. Aoki, S.-W. Lee, G. Chen, R. Wu, X. Pan, *Nature* **2022**, *606*, 292.

[29] E. Hoglund, H. Walker, M. Hussain, D.-L. Bao, H. Ni, A. Mamun, J. Baxter, A. Khan, S. Pantelides, P. Hopkins, J. Hachtel, **2023**, DOI 10.21203/rs.3.rs-3570025/v1.

[30] Peter Meisenheimer, Arundhati Ghosal, Eric Hoglund, Zhiyang Wang, Piush Behera, Fernando Gómez-Ortiz, Pravin Kavle, Evguenia Karapetrova, Pablo García-Fernández, L. W. Martin, Archana Raja, Long-Qing Chen, Patrick E. Hopkins, Javier Junquera, Ramamoorthy Ramesh, *Nano Lett.* **2024**, *24*, 2972.

[31] Z. Jiang, J. Wang, *Journal of Applied Physics* **2022**, *131*, 164101.

[32] C. Dai, Z. Hong, S. Das, Y.-L. Tang, L. W. Martin, R. Ramesh, L.-Q. Chen, *Applied Physics Letters* **2023**, *123*, 052903.

[33] P. Behera, M. A. May, F. Gómez-Ortiz, S. Susarla, S. Das, C. T. Nelson, L. Caretta, S.-L. Hsu, M. R. McCarter, B. H. Savitzky, E. S. Barnard, A. Raja, Z. Hong, P. García-Fernandez, S. W. Lovesey, G. van der Laan, P. Ercius, C. Ophus, L. W. Martin, J. Junquera, M. B. Raschke, R. Ramesh, *Science Advances* **2022**, *8*, eabj8030.

[34] K. X. Nguyen, Y. Jiang, M. C. Cao, P. Purohit, A. K. Yadav, J. Junquera, M. W. Tate, S. M. Gruner, R. Ramesh, S. Salahuddin, D. A. Muller, *Microsc Microanal* **2018**, *24*, 176.

[35] F. Gómez-Ortiz, P. García-Fernández, J. M. López, J. Junquera, *Phys. Rev. B* **2022**, *106*, 134106.

[36] L. Zhang, J. Han, H. Wang, W. A. Saidi, R. Car, W. E, **2018**, DOI 10.48550/arXiv.1805.09003.

[37] A. Hjorth Larsen, J. Jørgen Mortensen, J. Blomqvist, I. E. Castelli, R. Christensen, M. Dułak, J. Friis, M. N. Groves, B. Hammer, C. Hargus, E. D. Hermes, P. C. Jennings, P. Bjerre Jensen, J. Kermode, J. R. Kitchin, E. Leonhard Kolsbjerg, J. Kubal, K. Kaasbjerg, S. Lysgaard, J.

Bergmann Maronsson, T. Maxson, T. Olsen, L. Pastewka, A. Peterson, C. Rostgaard, J. Schiøtz, O. Schütt, M. Strange, K. S. Thygesen, T. Vegge, L. Vilhelmsen, M. Walter, Z. Zeng, K. W. Jacobsen, *J. Phys.: Condens. Matter* **2017**, *29*, 273002.

[38] T. W. Pfeifer, H. A. Walker, H. T. Aller, S. Graham, S. Pantelides, J. A. Hachtel, P. E. Hopkins, E. R. Hoglund, **2025**, DOI 10.48550/arXiv.2503.09792.

[39] J. Á. Castellanos-Reyes, P. M. Zeiger, J. Rusz, *Phys. Rev. Lett.* **2025**, *134*, 036402.

[40] A. Husain, M. Mitrano, M. Rak, P. Abbamonte, A. Kogar, S. Vig, L. Venema, V. Mishra, P. Johnson, G. Gu, E. Fradkin, M. Norman, *SciPost Physics* **2017**, *3*, 27.

# Acknowledgements

Vibrational EELS experiments were supported by the U.S. Department of Energy, Office of Basic Energy Sciences (DOE-BES), Division of Materials Sciences and Engineering under contract ERKCS89. 4D-STEM and mechanical cross-section sample preparation were performed as part of a user proposal at the Center for Nanophase Materials Sciences (CNMS), which is a US Department of Energy, Office of Science, User Facility. Microscopy performed using instrumentation within ORNL's Materials Characterization Core provided by UT-Battelle, LLC, under Contract No. DE-AC05- 00OR22725 with the DOE and sponsored by the Laboratory Directed Research and Development Program of Oak Ridge National Laboratory, managed by UT-Battelle, LLC, for the U.S. Department of Energy. T.R.L. and Y.T.S. acknowledge support from the U.S. Department of Energy, Basic Energy Sciences (award No. DE-SC0025423). The 4D-STEM data were acquired at the Core Center of Excellence in Nano Imaging at the University of Southern California. This work was supported by the Center for Quantum Sensing and Quantum Materials, an Energy Frontier Research Center funded by the U.S. Department of Energy (DOE), Office of Science, Basic Energy Sciences (BES), under Award No. DE-SC0021238. P.A. acknowledges additional support from the EPiQS program of the Gordon and Betty Moore Foundation under Grant No. GBMF9452.

# Author Contributions

E.R.H. was responsible for the concept of the work and managed the efforts. E.R.H., H.A.W., and J.A.H. were responsible for the writing and primary editing of the manuscript. E.R.H., and J.A.H performed vibrational EELS on the Nion HERMES. E.R.H performed 4D-STEM and imaging on the Nion U100 and Themis Z. A.R.L. and S.C.Q. provided support for all Nion instruments. T.L. and Y.S. were responsible for 4D-STEM on the Thermo Fisher Spectra. S.S. provided discussion during the beginning of experimentation. H.A.W., T.W.P., D.B., P.E.H., and S.T.P. were responsible for EELS and ML-MD simulations. P.M. and R.R. were responsible for growing the samples. N.D.V., D.C., and P.A. were responsible for the R-EELS.

# Competing Interests

S.C.Q. is an employee of Bruker.  The remaining Authors declare that they have no competing interests.

# Data Availability

The datasets generated during and/or analyzed during the current study are available from the corresponding authors on reasonable request.

## Extended Figures

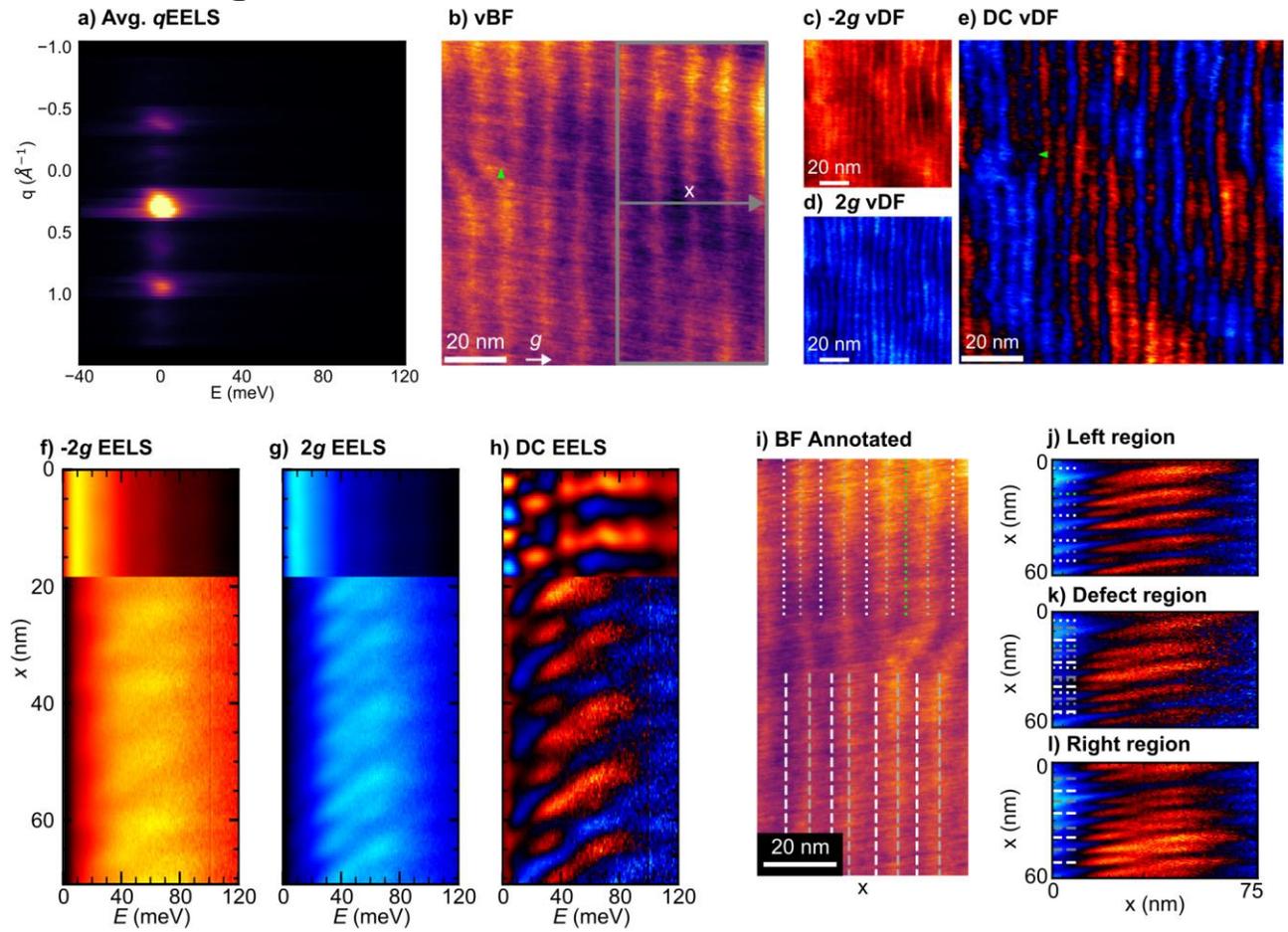

Figure E1. **Vibrational response from topological lattice defects.** (a) Spatially averaged *q*EELS that is dispersed in energy and shows the filtered diffraction peaks at zero-energy. (b) Virtual brightfield (vBF), and (c) -2g and (d) 2g virtual dark-field (vDF). (e) The zero-loss filtered vDF dichroic signal shows an effective polarization similar to Figure 1(d). (f) -2**g** and (g) +2**g** vDF EELS profiles taken perpendicular to the vortices in the non-defected region of (b). Their (h) dichroic signal showing spatially dependent dichroic vibrational loss at different regions following the vortex lattice periodicity. Left-hand simulated vibrational EELS line scans are overlayed on (f-g). (i) Enlarged and annotated region of the dislocation region in (b) with top (dotted) and bottom (dashed) bright (grey) and dark (white) vortex pairs labeled. The extra half plane of the dislocation is annotated in green. Line scan taken perpendicular to the vortices from the (j) top, (k) core, and (l) lower regions of (i) with the same vortex annotations.

# Supporting: Lattice Vibrations in a Chiral Polar Vortex lattice and at Topological defects

Eric R. Hoglund[1][*][‖], Harrison A. Walker[2,3][‖], Peter Meisenheimer[4], Thomas W. Pfeifer[1,5], Niels De Vries[6], Dipanjan Chaudhuri[6], Ting-Ran Liu[7], Steven C. Quillin[8], Sandhya Susarla[9], De-Liang Bao[2], Patrick E. Hopkins[5,10,11], Andrew Lupini[1], Yu-Tsun Shao[7,12], Peter Abbamonte[6], Ramamoorthy Ramesh[4,13], Sokrates T. Pantelides[2,3,14], and Jordan A. Hachtel[1][♠]

[1] Center for Nanophase Materials Sciences, Oak Ridge National Laboratory, Oak Ridge, TN 37830, USA

[2] Dept. of Physics and Astronomy, Vanderbilt University, Nashville, TN 37235, USA

[3] Interdisciplinary Materials Science Program, Vanderbilt University, Nashville, TN 37235

[4] Dept. of Materials Science and Engineering, University of California, Berkeley, CA 94720, USA

[5] Dept. of Mechanical and Aerospace Engineering, University of Virginia, Charlottesville, VA 22904, USA

[6] Materials Research Laboratory, University of Illinois, Urbana-Champaign, 104 S. Goodwin Ave., Urbana, IL, 61801, USA

[7] Mork Family Department of Chemical Engineering and Materials Science, University of Southern California, Los Angeles, CA, USA.

[8] Nion Co., 1102 8th St., Kirkland, WA 98033, USA

[9] Materials Science and Engineering, School for Engineering of Matter, Transport, and Energy, Arizona State University, Tempe, AZ, USA

[10] Dept. of Materials Science and Engineering, University of Virginia, Charlottesville, VA 22904, USA

[11] Dept. of Physics, University of Virginia, Charlottesville, VA 22904, USA

[12] Core Center of Excellence in Nano Imaging, University of Southern California, Los Angeles, CA, USA

[13] Dept. of Materials Science and Nanoengineering, Department of Physics and Astronomy, Rice University, Houston, TX, 77251 USA

[14] Dept. of Electrical and Computer Engineering, Vanderbilt University, Nashville TN 37235, USA

[‖]Contributed equally to the paper

[*]hoglunder@ornl.gov

[♠]hachtelja@ornl.gov

# Contents

# S1 Supporting Information: Figures

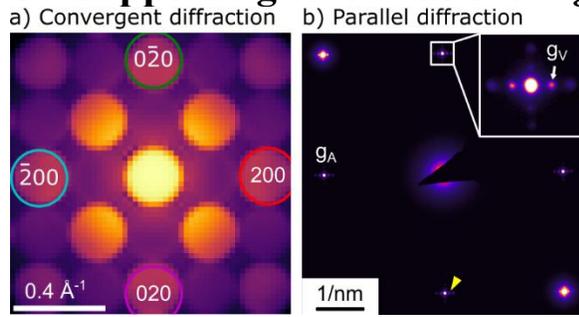

a) Convergent diffraction  b) Parallel diffraction

Figure S1. **Electron diffraction.** Diffraction patterns from a (a) convergent and (b) parallel probe illumination. Second order Friedel pairs from the atomic lattice that are used to form the center-of-mass dichroic signal in Figure 1(d) are labeled and annotated in (a). The inset in (b) shows an enlargement of one first order Bragg peaks ($g_A$) to emphasize satellite Bragg peaks ($g_V$) corresponding to diffraction the vortex lattice, which demonstrates long-range periodicity.

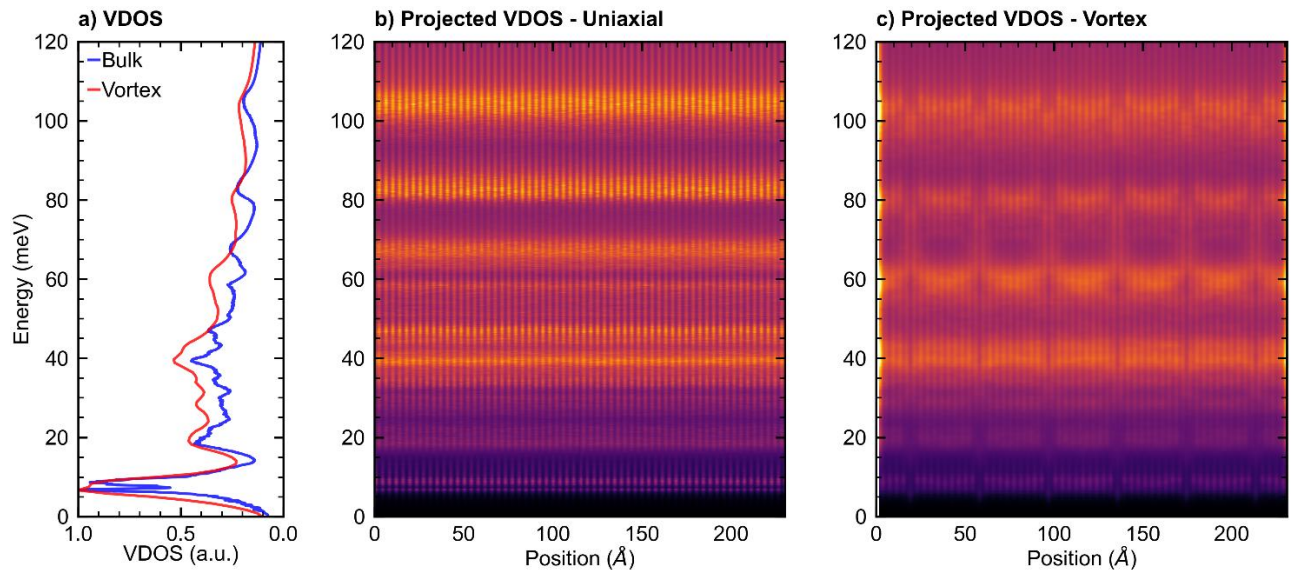

Figure S2. **Density of states.** (a) Total VDOS for the uniaxial and vortex supercells showing an apparent broadening and shifting of most peaks in the vortex cell data relative to the uniaxial. Full supercell of spatially projected VODS for the (b) uniaxial and (c) vortex supercells.

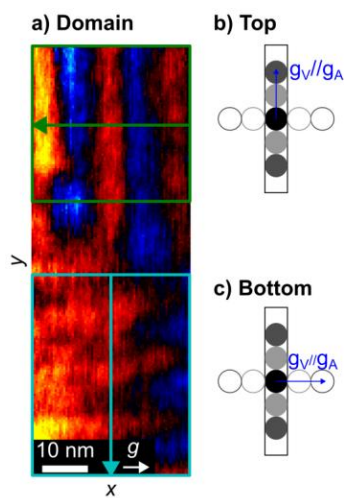

**a) Domain**   **b) Top**

$g_V//g_A$

**c) Bottom**

$g_V//g_A$

10 nm

*y*

*g*

*x*

Figure S3. **Vortex domain entrance aperture orientation.** (a) Dichroic image of the domain also shown in Figure 3(b). The white arrow indicates the long direction of the slot entrance aperture. Schematic of the entrance aperture orientation for the (b) top and (c) bottom domains. The shaded circles represent the Bragg peaks that are transmitted through the aperture. The blue arrow shows the direction of the vortex diffraction parallel to the atomic lattice, which is perpendicular to the vortices' axial direction. The entrance aperture collects the vortex diffraction in the top domain but not in the bottom domain.

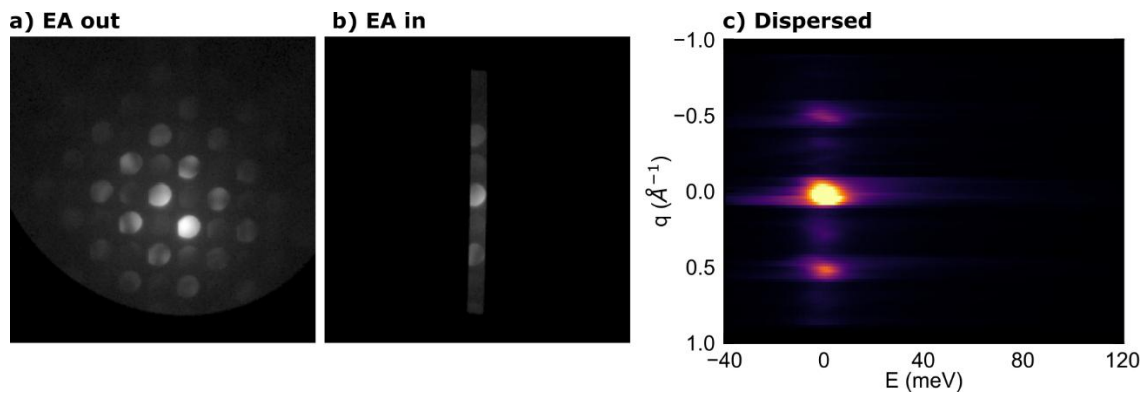

**a) EA out**   **b) EA in**   **c) Dispersed**

Figure S4. **Momentum resolved EELS experimental condition.** Diffraction pattern (a) without and (b) with the spectrometer entrance slot aperture. (b) Momentum resolved EELS spectrum.

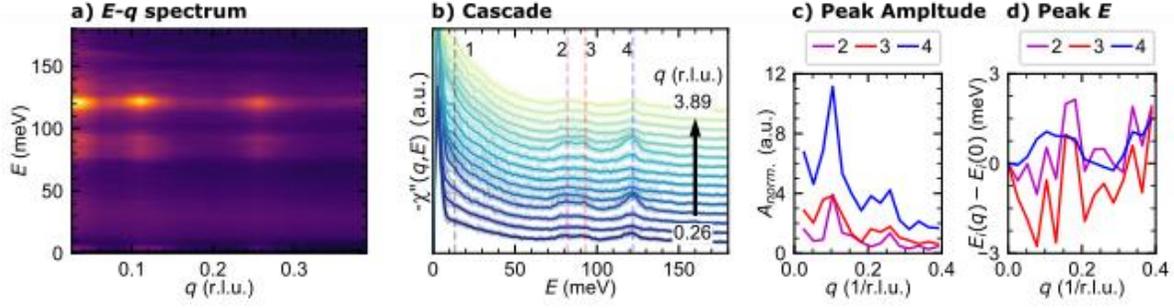

Figure S5. **R-EELS scans.** (a) Spectra stacked into an image and a (b) cascade. In (b) four primary peaks are marked by dashed lines and the three labeled ones were fit with Gaussians. The Gaussian amplitude is shown in (c). In (d) the difference Gaussian central energy relative to the energy at $q$=0 is plotted so that the oscillatory trend in energy can be observed independent of the absolute energy scales. It is worth noting that the energy separation of peaks 2 and 3 are less than the resolution times the dampening (line-width) so the peak fitting is prone to error, but general trends agree with peak 4, just with less statistical significance.

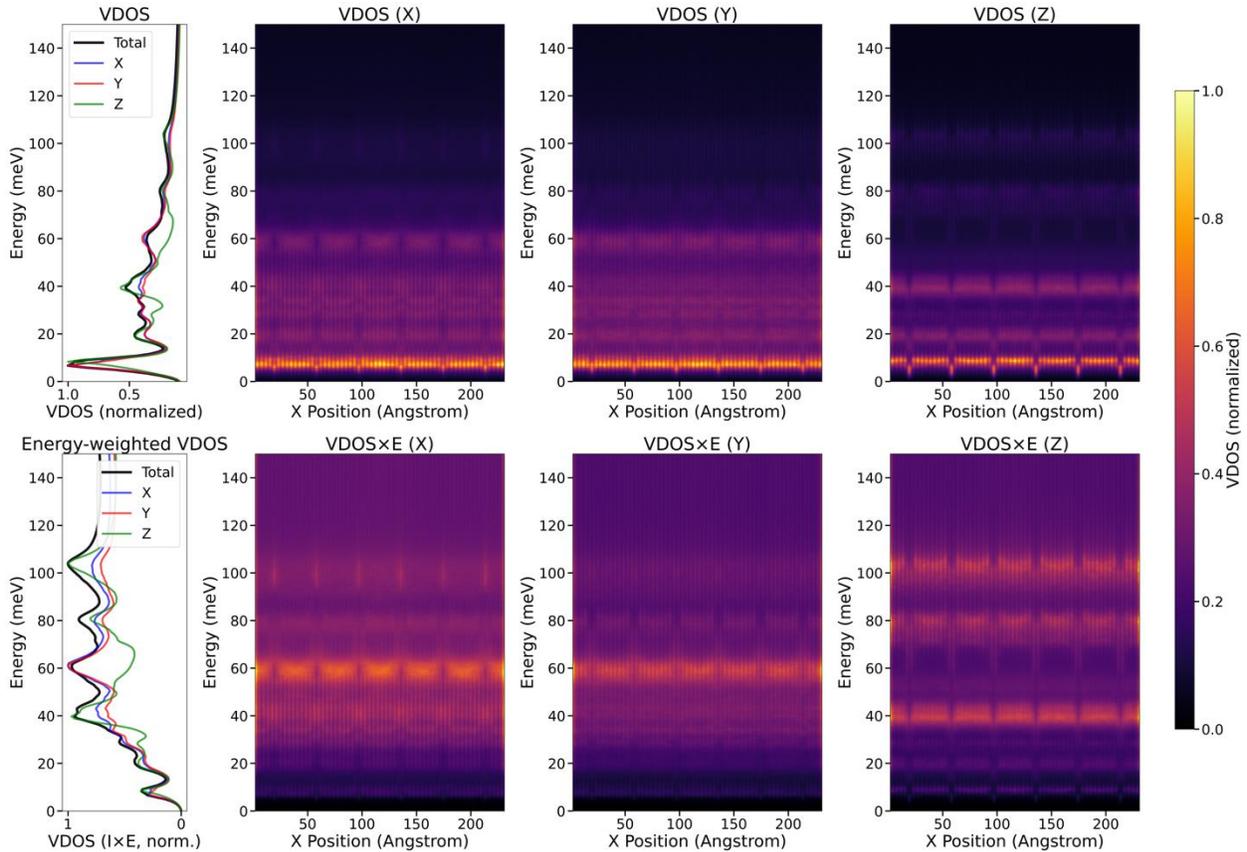

Figure S6. **Direction and Position Projected VDOS for the Vortex PbTiO₃.** (Top) DOS and (bottom) DOS with the I × E transform. From left to right column, VDOS, and x, y, z displacement projected VDOS.

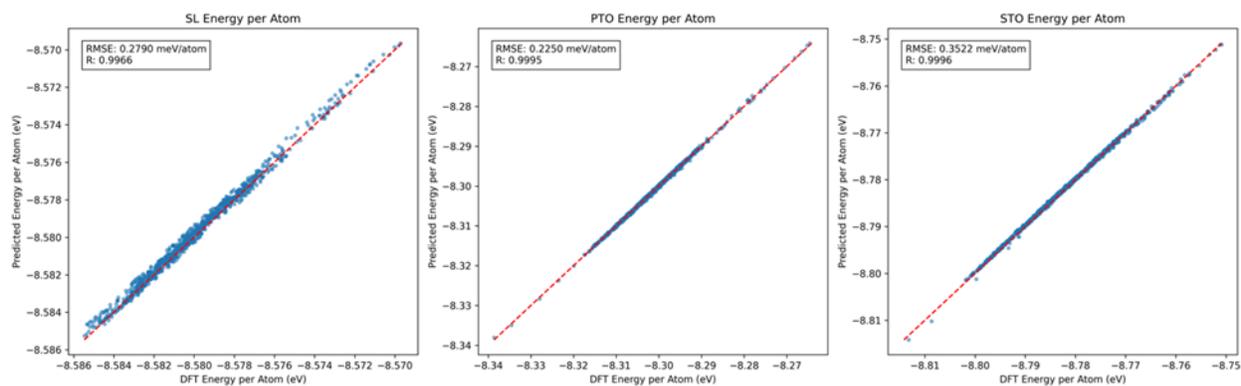

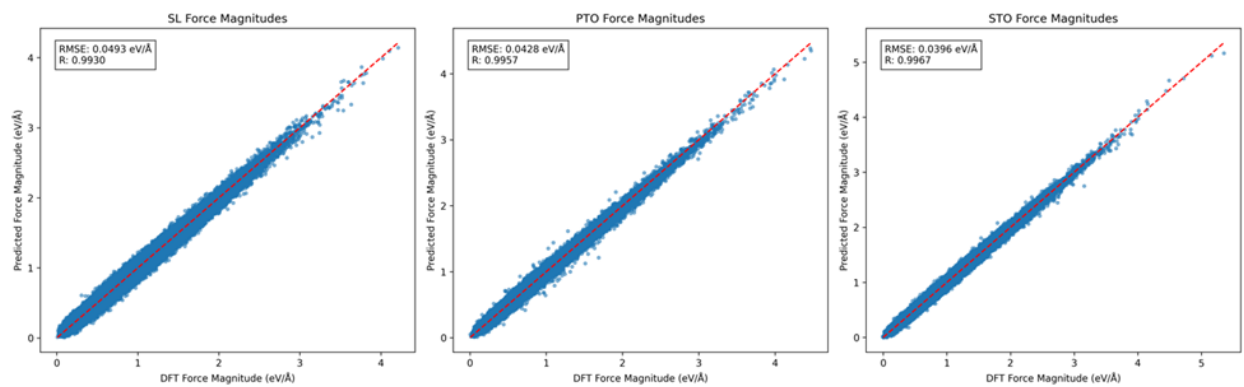

Figure S7. **Machine Learning Potential Parity Plots.** Each plot represents a subset of the test set encompassing bulk PbTiO$_3$, bulk SrTiO$_3$, and a superlattice of both at various temperatures. The plots compare predictions with the model to exact density functional theory calculations.